\documentclass[prl,twocolumn,floatfix,superscriptaddress, longbibliography]{revtex4-1}
\usepackage[pdftex,plainpages=false,colorlinks=true,linkcolor=blue, citecolor=blue, urlcolor=blue]{hyperref}
\usepackage{amsfonts}
\usepackage{amsmath}
\usepackage{amssymb}
\usepackage{graphicx}
\usepackage{natbib}
\usepackage{color}
\usepackage{placeins}
\usepackage{physics}

\begin{document}
 
\title{Entanglement entropy and mutual information across the Mott transition in the two-dimensional Hubbard model
}
\author{C. Walsh}
\affiliation{Department of Physics, Royal Holloway, University of London, Egham, Surrey, UK, TW20 0EX}
\author{P. S\'emon}
\affiliation{Computational Science Initiative, Brookhaven National Laboratory, Upton, NY 11973-5000, USA}
\author{D. Poulin}
\affiliation{D\'epartement de physique \& Institut quantique, Universit\'e de Sherbrooke, Sherbrooke, Qu\'ebec, Canada J1K 2R1}
\affiliation{Canadian Institute for Advanced Research, Toronto, Ontario, Canada, M5G 1Z8}
\author{G. Sordi}
\email[corresponding author: ]{giovanni.sordi@rhul.ac.uk}
\affiliation{Department of Physics, Royal Holloway, University of London, Egham, Surrey, UK, TW20 0EX}
\author{A.-M. S. Tremblay}
\affiliation{D\'epartement de physique \& Institut quantique, Universit\'e de Sherbrooke, Sherbrooke, Qu\'ebec, Canada J1K 2R1}
\affiliation{Canadian Institute for Advanced Research, Toronto, Ontario, Canada, M5G 1Z8}

\date{\today}

\begin{abstract}
Entanglement and information are powerful lenses to probe phases transitions in many-body systems. Motivated by recent cold atom experiments, which are now able to measure the corresponding information-theoretic quantities, we study the Mott transition in the half-filled two-dimensional Hubbard model using cellular dynamical mean-field theory, and focus on two key measures of quantum correlations: entanglement entropy and mutual information. We show that they detect the first-order nature of the transition, the universality class of the endpoint, and the crossover emanating from the endpoint.
\end{abstract}
 
\maketitle

{\it Introduction.--} 
The Mott transition is a metal-insulator transition caused by electron-electron interactions in a half-filled band~\cite{mott,ift}. As observed in a number of materials, it is a first-order transition that ends at finite temperature at a critical endpoint~\cite{ift, McWhan_Remeika_Rice_Brinkman_Maita_Menth_1971, limelette, Matsuura_Hiraka_Yamada_Endoh_2000,Kanoda:1997,Lefebvre:2000,PowellMcKenzieReview:2011,Furukawa_Kobashi_Kurosaki_Miyagawa_Kanoda_2018}. It plays a central role in the physics of quantum materials~\cite{ift, anderson:1987}. It is possible to explain this transition within the dynamical mean-field solution of the paradigmatic model of strongly correlated electrons, the Hubbard model~\cite{rmp}. That model can be mimicked accurately by ultracold atom experiments in optical lattices~\cite{Jordens:2008, Schneider:2008, Hofrichter:PRX2016, Cocchi:PRL2016}. 

Recent experimental progress with single-atom microscopy has bridged quantum materials and quantum information by measuring both the thermodynamics and the information-theoretic entanglement of the Hubbard model~\cite{greinerNat2015, Cocchi:PRX2017}. We take advantage of these recent experimental advances to open a new window on the interaction-driven Mott transition at half-filling.  

The Mott transition offers a unique opportunity to study information-theoretic measures of a first-order transition, from its quantum low temperature limit all the way to its finite-temperature critical endpoint and beyond. In the supercritical regime, crossover phenomena associated with the so-called Widom line~\cite{water1, supercritical, ssht} are expected. Information-theoretic measures have been used to probe quantum-critical points~\cite{Anfossi_Giorda_Montorsi_Traversa_2005,Amico_Patan_2007,amicoRMP2008,larssonPRA2006,LarssonScalingHubbard:2005} or finite-temperature continuous transition~\cite{Wilms_Troyer_Verstraete_2011} with the goal of understanding phases of matter as different structures of entanglement distribution in a system.

Motivated by recent experiments with ultracold atoms~\cite{Cocchi:PRX2017}, we characterise the interaction-driven Mott transition in the two-dimensional Hubbard model using quantum-information measures, as generalized to fermions by Zanardi~\cite{zanardi2002}. The entanglement entropy between a single site and the rest of the system emerged as a useful tool to characterise quantum phase transitions~\cite{amicoRMP2008}. At finite temperatures however, both entanglement and thermal fluctuations contribute to the single-site entropy $s_1$, which is therefore no longer a measure of quantum entanglement only~\cite{Cardy:2017}. The difference $\overline I_1 = s_1-s$ between the local entropy $s_1$ and the thermodynamic entropy $s$ leads to a measure of total mutual information that captures both classical and quantum correlations between a site and its environment~\cite{groisman2005}.  
We demonstrate that both $s_1$ and $\overline I_1$ (a) pinpoint the first-order nature of the Mott transition by showing hysteretic behavior; (b) detect the universality class of the Mott endpoint by showing critical scaling; (c) identify the crossover emanating from the endpoint in the supercritical region by showing sharp variations marked by inflections. 

Finite-temperature studies of information-theoretic measures of the Hubbard model have appeared, for example on the Kagome lattice~\cite{Udagawa_Motome:2015}, but previous studies of the Mott transition focused on bosonic systems~\cite{Lauchli_Kollath_2008, greinerNat2015, Frerot_Roscilde_2016} or on fermionic systems at zero temperature~\cite{Anfossi_Giorda_Montorsi_Traversa_2005, LarssonScalingHubbard:2005, larssonPRA2006, byczukPRL2012}.

{\it Model and method.--} 
We consider the single-band Hubbard model on the square lattice in two dimensions (2D) $H=-\sum_{\langle ij\rangle \sigma}t_{ij} c_{i\sigma}^\dagger c_{j\sigma}  +U\sum_{i} n_{i\uparrow } n_{i\downarrow }  -\mu\sum_{i\sigma} n_{i\sigma}, $ where $c^{\dagger}_{i\sigma}$ and $c_{i\sigma}$ operators create and destroy an electron of spin $\sigma$ on site $i$, $n_{i\sigma}=c^{\dagger}_{i\sigma}c_{i\sigma}$ is the number operator, $t_{ij}$ is the nearest neighbor hopping, $\mu$ is the chemical potential and $U$ the onsite Coulomb repulsion. 
We solve this model within plaquette cellular dynamical mean-field theory (CDMFT)~\cite{maier, kotliarRMP, tremblayR}, which is a cluster extension of dynamical mean-field theory (DMFT)~\cite{rmp}. 
The cluster in a bath problem is solved using the continuous-time quantum Monte Carlo method~\cite{millisRMP,patrickSkipList,patrickCritical} based on the hybridization expansion of the impurity action (CT-HYB). Further details are in the companion article~\cite{walshSb}. We work in units  $t=1$, $k_B=1$.

In the model that we study, the Mott transition is masked by long-range antiferromagnetic order. Nevertheless, introducing frustration decreases the antiferromagnetic transition temperature and can unmask the Mott transition~\cite{LorenzoAF, LorenzoAF2}. A model with frustrated antiferromagnetism would lead to increased fermionic sign problems, so we stick with the simpler model. While it will not be possible to compare quantitatively our low-temperature results with experiments where antiferromagnetism is frustrated, we do not expect qualitative differences. In calculations we do not allow antiferromagnetic symmetry breaking, so the Mott transition extends all the way to zero temperature.

\begin{figure}[t!]
\centering{
\includegraphics[width=1.\linewidth]{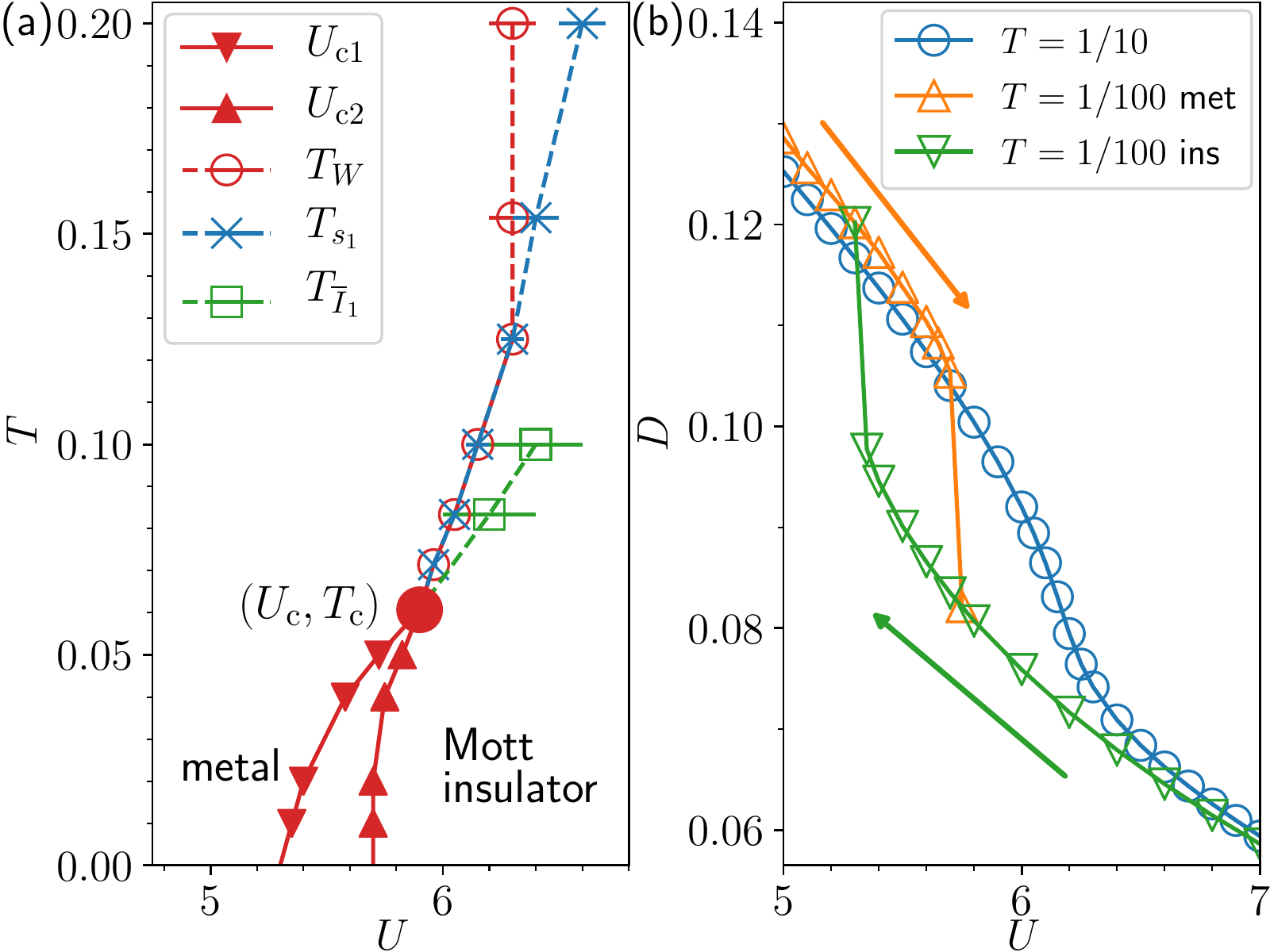}
}
\caption{(a) Temperature $T$ versus interaction strength $U$ phase diagram of the 2D Hubbard model at half filling ($n=1)$ within plaquette CDMFT. Coexistence associated with the first-order transition between a metal and a Mott insulator terminates at the second-order critical endpoint $(U_{\rm c}, T_{\rm c})$. From the endpoint emerges a crossover line (Widom line) defined by the maxima in the correlation length. Here we estimate the Widom line by the locus of the inflection point in the double occupancy $D(U)_T$ at different temperatures (open red circles, $T_W$). This line overlaps with the loci of the inflection points in the local entropy $s_1(U)_T$ (blue crosses, $T_{s_1}$) and in the total mutual information $\overline I_1(U)_T =s_1-s$ (green squares, $T_{\overline I_1}$), where $s$ is thermal entropy. (b) Double occupancy $D$ versus $U$ at $n=1$ and different temperatures. For $T<T_{\rm c}$, hysteresis appears when sweeping $U$ in the forward and reverse directions. Arrows show the sweep direction. 
}
\label{fig1}
\end{figure}

{\it Phase diagram.--}
The temperature versus interaction strength phase diagram of the half-filled 2D Hubbard model obtained by CDMFT on a 2x2 plaquette is known~\cite{phk, balzer, sht2}. As shown in Fig.~\ref{fig1}(a), at low $T$ and finite $U$, the electron-electron correlations produce a first-order transition between a metal and a Mott insulator. This transition terminates in a second-order critical endpoint at $(U_{\rm c}, T_{\rm c})$. From the endpoint emanates a crossover line, the Widom line, across which the correlation length peaks~\cite{water1, supercritical, ssht}. Near the endpoint, the loci of other response functions extrema converge into the Widom line~\cite{water1, supercritical}. Hence, the Widom line is a high-temperature precursor of the first-order Mott transition. 

A convenient way to construct the $T-U$ phase diagram of the half-filled model is to monitor the behavior of double occupancy $D(U)_T$ [see Fig.~\ref{fig1}(b)]. For $T<T_{\rm c}$, hysteresis loops reveal the coexistence region $U_{c1}(T)<U<U_{c2}(T)$ between metal and insulator. 
At $T=T_{\rm c}$ the coexistence region closes, $D(U)$ is continuous with an infinite slope at $U_{\rm c}$, and thermodynamic quantities such as double-occupancy fluctuations $\partial D/\partial U$ diverge, thereby signalling a diverging correlation length. For $T>T_{\rm c}$, $D(U)$ still shows an inflection point.
We use the inflection point in $D(U)_T$ to estimate the Widom line (see also companion article~\cite{walshSb}). 

The important aspect is that thermodynamic properties, from low temperature up to high temperature, are controlled by the transition and its associated supercritical crossover. Here we shall extend this connection to information-theoretic measures.

{\it Local entropy and entanglement entropy --.} 
The entropy of a subsystem $A$ is defined as $s_A=- {\rm Tr_A} [\rho_A \ln \rho_A]$, where the reduced density matrix $\rho_A$ is obtained by tracing the complement $B$ of $A$ from the global density matrix, $\rho_A = {\rm Tr}_B [\rho_{AB}]$. 
It is a measure of the lack of information, or uncertainty, in the state of $A$. At $T=0$, this uncertainly is due to the entanglement between $A$ and $B$, so $s_A$ is called entanglement entropy. Here we focus on the local entanglement entropy~\cite{zanardi2002,larssonPRA2006}, where $A$ is a site of the lattice and $B$ the remaining sites. The state space of a single site is spanned by $\{ \ket{0}, \ket{\uparrow}, \ket{\downarrow}, \ket{\uparrow\downarrow} \}$. An important simplification occurs due to spin conservation, which ensures that the reduced density matrix is  diagonal~\cite{zanardi2002}, $\rho={\rm diag}(p_0, p_\uparrow, p_\downarrow, p_{\uparrow \downarrow})$, where $p_i$, with  $i=\{ 0, \uparrow, \downarrow, \uparrow\downarrow \}$, is the probability for a site to be empty, occupied with a spin up or down particle or doubly occupied. One has $p_{\uparrow \downarrow}=\langle n_{i\uparrow} n_{i\downarrow} \rangle=D$, $p_{ \uparrow} = p_{\downarrow} = \langle n_{i\uparrow} -n_{i\uparrow} n_{i\downarrow} \rangle$ and $p_{0}=1-2p_\uparrow -p_{\uparrow\downarrow}$. Thus $s_1$ becomes $s_1 = -\sum_i p_i \ln(p_i),$ that can be easily calculated with CDMFT and measured in ultracold atom experiments~\cite{Cocchi:PRX2017}. 
For Hubbard-like Hamiltonians, $s_1$ emerged as a useful tool to detect phase transitions and critical behavior, especially at $T=0$ where it measures entanglement~\cite{amicoRMP2008, Gu:2004, LarssonScalingHubbard:2005, larssonPRA2006, CamposPRA2006, byczukPRL2012}. 
Our first contribution is to extend these studies to the first-order Mott transition from its low temperature quantum regime to the critical endpoint and supercritical crossovers, where $s_1$ acquires thermal contributions and is no longer a measure of quantum entanglement only~\cite{Cardy:2017}.

\begin{figure}[t!]
\centering{
\includegraphics[width=1\linewidth]{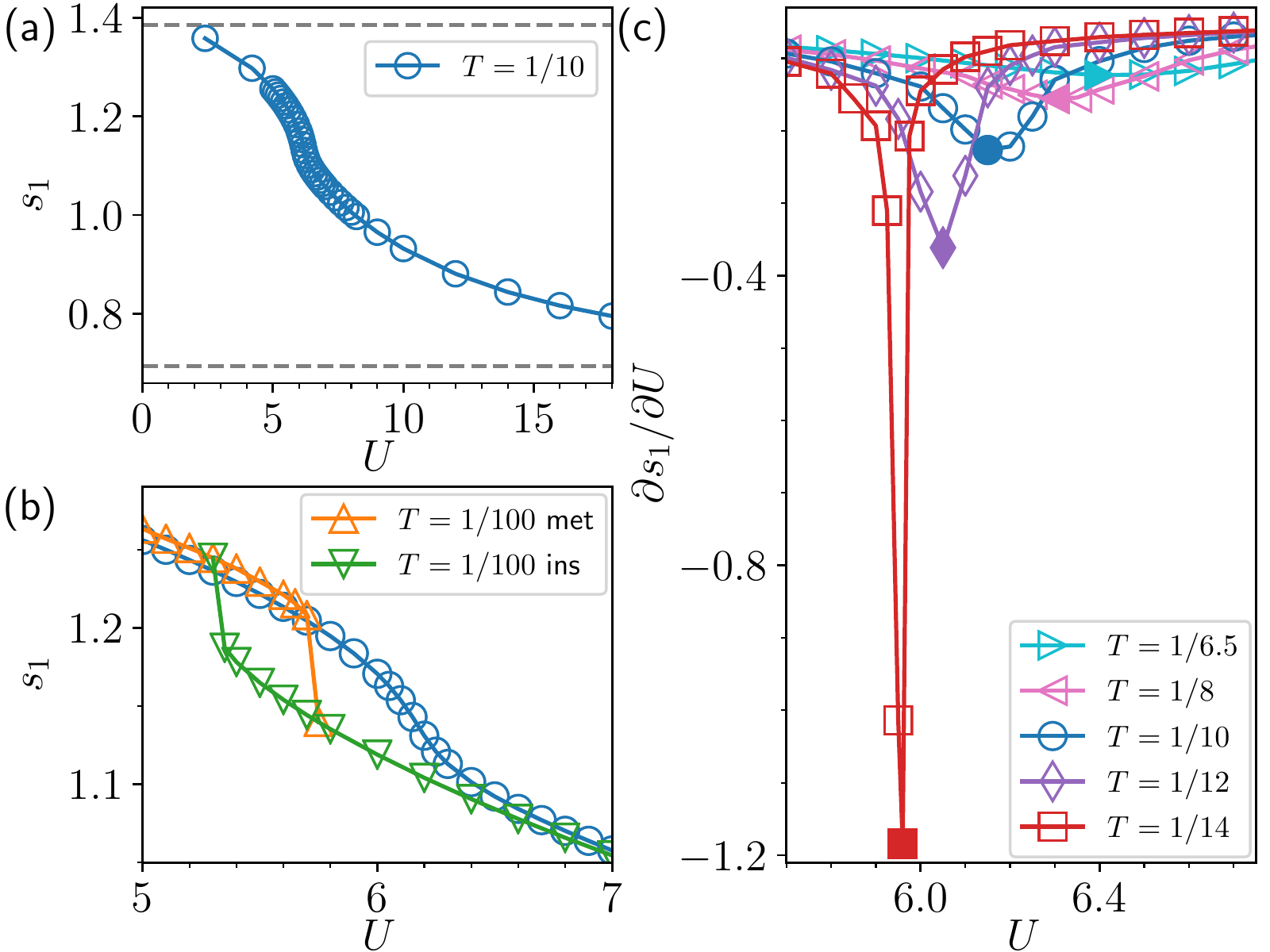}
}
\caption{(a) Local  entropy $s_1$ versus $U$ at $n=1$ and $T=1/10$ (circles). Horizontal lines mark $\ln 4$ (for $U=0$) and $\ln 2$ (for $U=\infty$). (b) Zoom-in of $s_1$ versus $U$ at $n=1$ for $T=1/10$ and $T=1/100$ (triangles) close to the Mott transition. Inflection point for $T>T_{\rm c}$ and jumps for $T<T_{\rm c}$ are visible. (c) Increase in the slope $\partial s_1/\partial U$ as one approaches $T_{\rm c}$ from above. The positions of the minima are shown as blue crosses in Fig.~\ref{fig1}(a).
}
\label{fig2}
\end{figure}

The local  entropy $s_1$ is shown as a function of $U$ and different temperatures in Fig.~\ref{fig2}(a),(b). 
$s_1(U)$ monotonically decreases from a maximum value of $\ln 4$ at $U=0$ to asymptotically reach the minimum value of $\ln 2$ for $U\rightarrow \infty$. In these limits, $\rho_A$ has equal populations in, respectively, four and two states. Physically, $U$ suppresses the charge fluctuations and therefore the number of available states. In information theory language, a larger $U$ leads to more knowledge about the site occupation, and hence to a decrease of its entropy.

Contrary to $s$, which is constrained to obey $\partial s/\partial T >0$ for reasons of thermodynamic stability, $\partial s_1/\partial T < 0$ is allowed. As can be seen by comparing the results for $T=1/100$ and $T=1/10$ in Fig.~\ref{fig2}(b), this occurs on the metallic side because, at fixed filling, the number of available states on a single site can decrease when $D(T)$ decreases with increasing $T$, a phenomenon which is known to occur at low $T$~\cite{rmp, WernerAdiabCooling2005, Dare:2007, Paiva_Scalettar_Randeria_Trivedi_2010}. 
For $T<T_{\rm c}$, $s_1(U)$ shows hysteresis loops at the Mott transition [see Fig.~\ref{fig2}(b)], with $(s_{1})_{\rm ins} < (s_{1})_{\rm met}$ since double occupancy is smaller in the insulating state. 
For $T\ge T_{\rm c}$, $s_1(U)_T$ shows an inflection point. Approaching $T_{\rm c}$ from above, the first derivative $\partial s_1/ \partial U$ becomes more negative and eventually tends to minus infinity at the Mott endpoint (see Fig.~\ref{fig2}(c)). This follows from $(\partial s_1/\partial D) (\partial D /\partial U)$ with $\partial s_1/\partial D$ regular and $\partial D / \partial U \sim -|U-U_{\rm c}|^{-1 +1/\delta}$ with $\delta>1$~\cite{patrickCritical}. The locus of the inflection point of $s(U)_T$ at different temperatures is shown in Fig.~\ref{fig1}(a) and defines a crossover which is a precursor of the Mott transition in the supercritical region. Moreover, the asymptotic behavior of $s_1(U)$ at $T_{\rm c}$ allows us to extract directly the critical exponent $\delta$. Although critical exponents are not generally observable in $s_1$~\cite{larssonPRA2006}, here this is possible since $s_1$ depends only on double-occupancy. 
The loci of the inflections of $s_1(U)_T$ and of $D(U)_T$ do not need to coincide in the supercritical region, they only need to converge towards each other at the critical endpoint. Surprisingly, inspection of Fig.~\ref{fig1}(a) shows that the loci of inflections in $s_1(U)_T$ and $D(U)_T$ remain close, up to high temperature, suggesting an interesting connection between fluctuations of the double occupancy, $\partial D/\partial U$, and variations in the entanglement entropy, $\partial s_1/\partial U$~\cite{LarssonScalingHubbard:2005}. 
Overall, these findings provide a strong link between thermodynamics and entanglement: the Mott transition, critical scaling, and associated high-temperature crossovers, can be determined solely from the local entropy $s_1$, without knowledge of the order parameter of the transition.

{\it Total mutual information.--}
The mutual information $I(A:B)$ captures the total correlations between two systems $A$ and $B$. In particular, it takes value 0 if and only if the two systems are uncorrelated $\rho_{AB} = \rho_A\otimes \rho_B$. It is defined as $I(A:B) = s_A+s_B-s_{AB}$, and appears for instance in classical and quantum information theory~\cite{CoverInformation, watrous2018}. 

The mutual information $I_n$  between a ball $A$ of volume $n$ and the rest of the lattice $B$ is the focus of the famous area law \cite{EisertArea2010}. At finite temperature, it is known~\cite{wolfPRL2008} that for a system of volume $n$, this quantity scales at most proportionally to $n^{\frac{d-1}d}$, the size of the boundary separating $A$ and $B$. This is attributed to the fact that sites deep in the bulk --- sites whose distance from the $A/B$ boundary is larger than the thermal correlation length --- do not contribute to the mutual information. This scaling is conjectured to remain true at $T=0$ when the system has a spectral gap $\Delta>0$, and the argument simply replaces the thermal correlation length by the quantum correlation length $\xi \propto 1/\Delta$. However, this prediction has only been proven rigorously in 1D~\cite{Hasting2007, Aharonov2011, Arad2012, Arad2013}.  

Here, we are interested in the mutual information between a given site $i$ and the rest of the lattice, averaged over all sites. For the site labeled $i=1$, we have $I(1:\{>1\}) = s_1+s_{\{>1\}} - s_{\{>0\}}$ where we denote by $\{>k\}$ the set of sites with indices greater than $k$, so $\{>0\}$ is the entire lattice. For the site labeled $i=2$, the mutual information between $i=2$ and the rest of the lattice $\{1\}\cup \{>2\}$ would lead to double counting of the correlations between site 1 and 2, that have already been account for in the quantity $I(1:\{>1\})$. To avoid such double-counting, we trace over site 1, that has already been considered, and compute the mutual information between site $2$ and the remaining sites $\{>2\}$. Continuing this process, we define the total mutual information between a single site and the rest of the lattice as
\begin{equation*}
\overline I_1 =  \frac 1N \sum_{i=1}^N I(i:\{>i\}) = \frac 1N \sum_{i=1}^N (s_i + s_{\{>i\}} - s_{\{>i-1\}}).
\end{equation*}
It is easy to see from this last form that most terms cancel, leaving $\overline I_1 =  ( \sum_{i=1}^N s_i/N - s)$ where $s$ is the thermodynamic entropy, defined as the entropy of the entire lattice divided by the number of sites. For a translationally invariant system, all $s_i$ are equal, so the total mutual information further simplifies to the difference between the local entropy and the thermodynamic entropy $\overline I_1 = s_1-s$.

\begin{figure}[t!]
\centering{
\includegraphics[width=1.\linewidth]{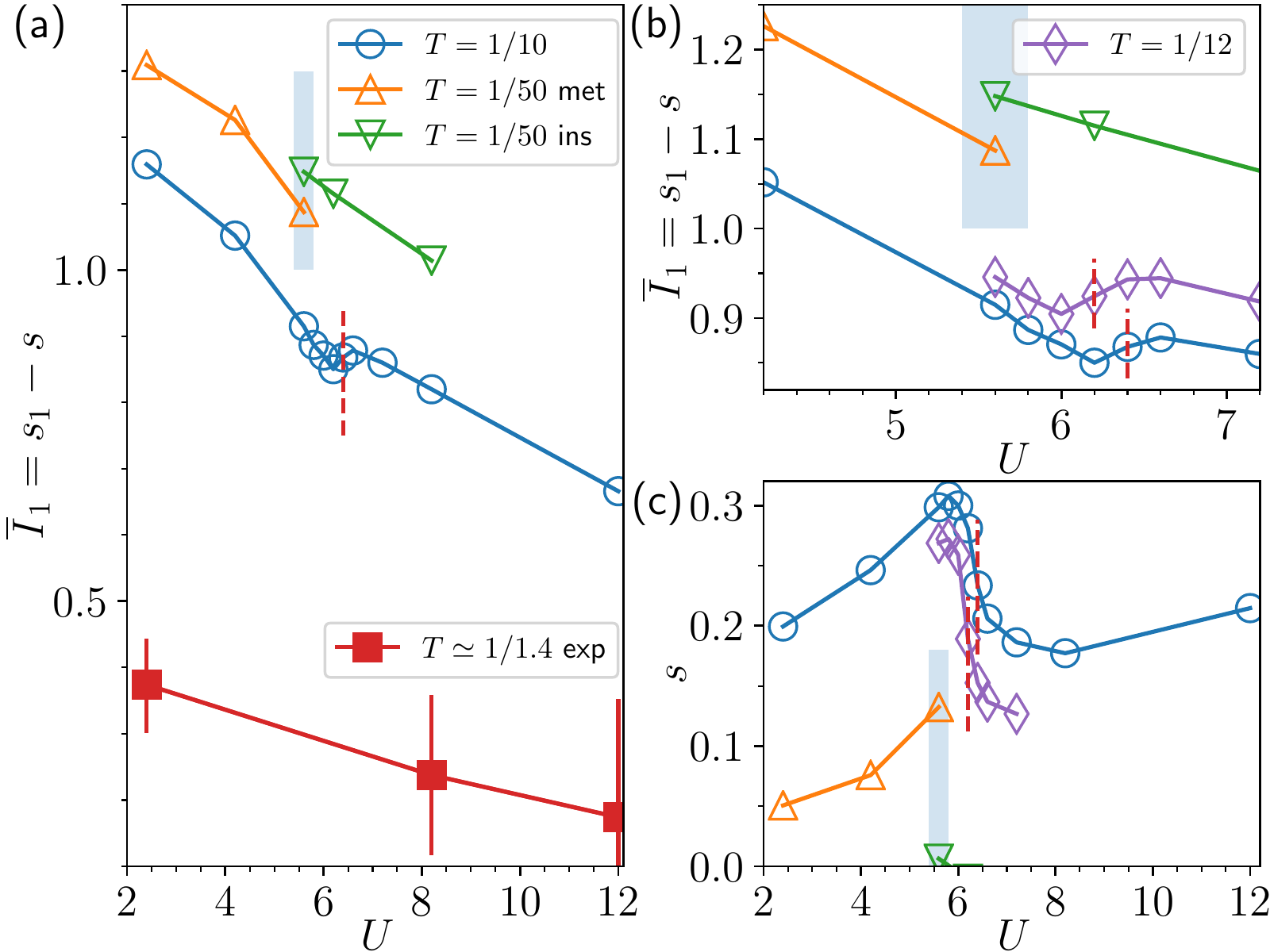}
}
\caption{(a) Total mutual information $\overline I_1$ versus $U$ at $n=1$ for different temperatures. Filled squares are experimental data of ultracold atoms in Ref.~\cite{Cocchi:PRX2017} at $T\approx 1/1.4$. (b) Zoom-in of panel (a) showing $T=1/10, 1/12$, and $T=1/50$. (c) Thermodynamic entropy $s$ versus $U$ at $n=1$ for the same temperature of panel (b). In all panels, a dashed vertical line marks the inflection point while the shaded area marks the coexistence between metal and insulator. The loci of the inflections in $\overline I_1(U)_T$ are shown as green squares in Fig.~\ref{fig1}(a). 
}
\label{fig3}
\end{figure}

Fig.~\ref{fig3}(a) shows $\overline I_1$ as a function of $U$ and different temperatures (open symbols). 
The thermodynamic entropy $s(U)$ is shown in Fig.~\ref{fig3}(c) and discussed in the companion article~\cite{walshSb}.  
Far from the Mott transition, Fig.~\ref{fig3}(a) for $\overline I_1$ shows that we find qualitative consistency with experimental data at higher temperature (filled squares): $\overline I_1$ at weak interaction is larger than $\overline I_1$ at strong interaction, because for weaker interaction the states are more extended, hence the density matrix does not factor in position space. 
Nevertheless $\overline I_1$ can be quite large in the Mott insulator. At finite $U$ electrons are localised, but their spins lock into singlet states due to the superexchange mechanism~\cite{phk, sht, sht2}. Hence they are correlated and $\overline I_1$ cannot be zero. That is a basic message of the CDMFT: localisation along with short-range correlations are key ingredients of the Mott transition, and this remarkably shows up in $\overline I_1$. 

Consider now the vicinity of the Mott transition. For $T<T_{\rm c}$, $\overline I_1$ shows hysteresis [triangles in Fig.~\ref{fig3}(a)]. Although $s_1$ is larger in the metal than in the insulator [Fig.~\ref{fig2}(b)], the entropy $s$ of the insulator is much smaller than that of the metal [Fig.~\ref{fig3}(c)], leading to $(\overline I_1)_{\rm met} < (\overline I_1)_{\rm ins}$ across the Mott transition. The latter inequality seems counterintuitive, but may be understood as follows: while local charge correlations are stronger in a delocalised conducting state, a localised insulating state leads to a stronger superexchange coupling, which results in increased spin correlations that overwhelm the loss of charge correlations~\cite{phk, sht, sht2}.
For $T\ge T_{\rm c}$ [circles and diamonds in Figs.~\ref{fig3}(a),(b)], $\overline I_1$ displays non-monotonic behavior: there is a local minimum that comes from the displacement of the local minimum in $-s$ [Fig.~\ref{fig3}(c)] caused by the monotonic decrease of $s_1(U)$. The rapid increase beyond the minimum reflects the steep slopes in $s(U)$ and $s_1(U)$, marked by inflections [vertical red dashed lines], that eventually become infinite at $(U_{\rm c},T_{\rm c})$ since both quantities scale as $-{\rm sgn}(U-U_{\rm c}) |U-U_{\rm c}|^{1/\delta}$. 
For the Ising universality class, to which the Mott transition belongs, $\overline I_1$ scales the same way $\overline I_1 \sim {\rm sgn}(U-U_{\rm c}) |U-U_{\rm c}|^{1/\delta}$. 
The locus of the inflection marking the most rapid {\it increase} of $\overline I_1(U)_T$ at different temperatures defines the crossover line $T_{\overline I_1}$ in the $T-U$ phase diagram [see squares in Fig.~\ref{fig1}(a)]. 
Therefore, $\overline I_1$ detects the Mott transition, its supercritical crossover and critical behavior at the Mott endpoint.

{\it Conclusion.--} 
We have characterized the Mott transition in the 2D Hubbard model with CDMFT using information-theoretic measures. The local entropy $s_1$ and the total mutual information $\overline I_1$ both detect the first-order nature of the transition, the criticality of the Mott endpoint and the supercritical crossover. 
Our results are consistent with ultracold atom experiments of Ref.~\cite{Cocchi:PRX2017} and provide specific predictions for information-theoretic measures in the low-temperature region that may soon be accessible. 

In highly frustrated optical lattices, such as the triangular lattice, one can in principle measure $s_1$ and $\overline I_1$ across the Mott transition and its supercritical crossover. One could thus verify our results of sharp variations of these quantities as a function of $U$, characterized by inflection points (above $T_{\rm c}$) and by hysteresis (below $T_{\rm c}$). 

At the theory level, our information-theoretic description of a first-order transition from its quantum low temperature limit ($T \ll T_{\rm c}$), to its finite-temperature critical endpoint (at $T_{\rm c}$) and further on in the supercritical region ($T>T_{\rm c}$) offers a new kind of `criticality'~\cite{sht, sht2}, where strong variation of entanglement properties are not associated with a quantum critical point between ordered phases, but by proximity to a critical endpoint. 
Our finding of a deep link between the behavior of information-theoretic measures and of thermodynamic quantities can be extended to other first-order transitions ending in a critical endpoint, such as those found in electron systems~\cite{sht, sht2, ssht, walsh2}, in spin systems~\cite{Wilms_Troyer_Verstraete_2011, Iaconis:2013}, and quark matter~\cite{eeQCD2017}.

\begin{acknowledgments}
We acknowledge Janet Anders and Marcelo Rozenberg for useful discussions. This work has been supported by the Natural Sciences and Engineering Research Council of Canada (NSERC) under grants RGPIN-2014-04584, RGPIN-2014-06630, the Canada First Research Excellence Fund and by the Research Chair in the Theory of Quantum Materials. Simulations were performed on computers provided by the Canadian Foundation for Innovation, the Minist\`ere de l'\'Education des Loisirs et du Sport (Qu\'ebec), Calcul Qu\'ebec, and Compute Canada.
\end{acknowledgments}


%

\end{document}